\documentclass[aps,prl,showpacs,superscriptaddress,amssymb,nofootinbib,twocolumn]{revtex4}

\usepackage{graphicx}
\usepackage{epsfig}

\begin{document}

\title{Dirac Equation and Quantum Relativistic Effects in a Single Trapped Ion}

\author{L. Lamata}
\affiliation{Instituto de Matem\'aticas y F\'{\i}sica Fundamental,
CSIC, Serrano 113-bis, 28006 Madrid, Spain}
\author{J. Le\'on}
\affiliation{Instituto de Matem\'aticas y F\'{\i}sica Fundamental,
CSIC, Serrano 113-bis, 28006 Madrid, Spain}
\author{T. Sch\"atz}
\affiliation{Max-Planck-Institut f\"{u}r Quantenoptik,
Hans-Kopfermann-Strasse 1, D-85748 Garching, Germany}
\author{E. Solano}
\affiliation{Physics Department, ASC, and CeNS,
Ludwig-Maximilians-Universit\"at, Theresienstrasse 37, 80333
Munich, Germany } \affiliation{Secci\'{o}n F\'{\i}sica,
Departamento de Ciencias, Pontificia Universidad Cat\'{o}lica del
Per\'{u}, Apartado Postal 1761, Lima, Peru}

\date{\today}

\begin{abstract}
We present a method of simulating the Dirac equation in 3+1
dimensions for a free spin-$1/2$ particle in a single trapped ion.
The Dirac bispinor is represented by four ionic internal states,
and position and momentum of the Dirac particle are associated
with the respective ionic variables. We show also how to simulate
the simplified 1+1 case, requiring the manipulation of only two
internal levels and one motional degree of freedom. Moreover, we
study relevant quantum-relativistic effects, like the {\it
Zitterbewegung} and Klein's paradox, the transition from massless
to massive fermions, and the relativistic and nonrelativistic limits,
via the tuning of controllable experimental parameters.
\end{abstract}


\maketitle

The search for a fully relativistic Schr\"{o}dinger equation gave
rise to the Klein-Gordon and Dirac equations. P. A.
M. Dirac looked for a Lorentz-covariant wave equation that is
linear in spatial and time derivatives, expecting that the
interpretation of the square wave function as a probability
density holds. As a result, he obtained a fully covariant wave
equation for a spin-$1/2$ massive particle, which
incorporated {\it ab initio} the spin degree of freedom. It is
known~\cite{Thaller} that the Dirac formalism describes accurately the spectrum
of the hydrogen atom and that it plays a central role in quantum
field theory, where creation and annihilation of particles are
allowed. However, the one-particle solutions of the Dirac equation
in relativistic quantum mechanics predict some astonishing
effects, like the {\it Zitterbewegung} and the Klein's
paradox.

In recent years, a growing interest has appeared regarding
simulations of relativistic effects in controllable physical
systems. Some examples are the simulation of Unruh effect in
trapped ions~\cite{Alsing1}, the {\it Zitterbewegung} for massive
fermions in solid state physics~\cite{Schliemann}, and black-hole
properties in the realm of Bose-Einstein condensates~\cite{Garay}.
Moreover, the low-energy excitations of a
nonrelativistic two-dimensional electron system in a single layer
of graphite (graphene) are known to follow the Dirac-Weyl
equations for massless relativistic
particles~\cite{Cserti,Katsnelson}. On the other hand, the fresh dialog
between quantum information and special relativity has raised
important issues concerning the quantum information content of
Dirac bispinors under Lorentz transformations~\cite{PeresReview}.

In this Letter, we propose the simulation of the Dirac equation
for a free spin-$1/2$ particle in a single trapped ion. We show
how to implement realistic interactions on four ionic internal
levels, coupled to the motional degrees of freedom, so as to
reproduce this fundamental quantum relativistic wave equation. We
propose also the simulation of the Dirac equation in 1+1
dimensions, requiring only the control of two internal levels and
one motional degree of freedom. We study some
quantum-relativistic effects, like the {\it Zitterbewegung} and
the Klein's paradox, in terms of measurable observables. Moreover, we discuss the transition from massless
to massive fermions, and from the relativistic to the nonrelativistic limit. Finally, we describe a possible experimental scenario.

We consider a single ion of mass $M$ inside a Paul trap with
frequencies $\nu_x$, $\nu_y$, and $\nu_z$, where four metastable
ionic internal states, $| a \rangle, | b \rangle, | c \rangle$,
and $| d \rangle$, may be coupled pairwise to the center-of-mass
(CM) motion in directions $x$, $y$, and $z$. We will make use of
three standard interactions in trapped-ion technology, allowing
for the coherent control of the vibronic
dynamics~\cite{NIST-Innsbruck}. First, a carrier interaction
consisting of a coherent driving field acting resonantly on a pair
of internal levels, while leaving untouched the motional degrees
of freedom. It can be described effectively by the Hamiltonian
$H_{\sigma} = \hbar \Omega ( \sigma^{+} e^{i \phi} + \sigma^{-}
e^{-i\phi})$, where $\sigma^{+}$ and $\sigma^{-}$ are the raising
and lowering ionic spin-$1/2$ operators, respectively, and
$\Omega$ is the associated coupling strength. The phases and
frequencies of the laser field could be adjusted so as to produce
$H_{\sigma_x} = \hbar \Omega_x \sigma_x$, $H_{\sigma_y} = \hbar
\Omega_y \sigma_y$, and $H_{\sigma_z} = \hbar \Omega_z \sigma_z$,
where $\sigma_x$, $\sigma_y$, and $\sigma_z$, are atomic Pauli
operators in the conventional directions $x$, $y$, and $z$.
Second, a Jaynes-Cummings (JC) interaction, usually called
red-sideband excitation, consisting of a laser field acting
resonantly on two internal levels and one of the vibrational CM
modes. Typically, a resonant JC coupling induces an excitation in
the internal levels while producing a deexcitation of the motional
harmonic oscillator, and viceversa. The resonant JC Hamiltonian
can be written as $H_{\rm r} = \hbar \eta \tilde\Omega (\sigma^{+}
a e^{i \phi_{\rm r}} + \sigma^{-} a^{\dagger} e^{-i \phi_{\rm
r}})$, where $a$ and $a^{\dagger}$ are the annihilation and
creation operators associated with a motional degree of freedom.
$\eta = k \sqrt{\hbar / 2 M \nu}$ is the Lamb-Dicke
parameter~\cite{NIST-Innsbruck}, where $k$ is the wave number of
the driving field. Third, an anti-JC (AJC) interaction, consisting
of a JC-like coupling tuned to the blue motional sideband with
Hamiltonian $H_{\rm b} = \hbar \eta \tilde\Omega (\sigma^{+}
a^{\dagger} e^{i \phi_{\rm b}} + \sigma^{-} a e^{-i \phi_{\rm
b}})$. In this case, an internal level excitation accompanies an
excitation in the considered motional degree of freedom, and
viceversa.

All these interactions could be applied
simultaneously and addressed to different pairs of internal levels
coupled to different CM modes. For example, it is possible to
adjust field phases to implement a simultaneous blue and red
sideband excitation scheme to form the Hamiltonian
$H^{p_x}_{\sigma_x} = i \hbar \eta_x \tilde\Omega_x \sigma_x
(a^{\dagger}_x - a_x)  = 2  \eta_x \Delta_x \tilde\Omega_x
\sigma_x p_x$, with $i (a_x^{\dagger} - a_x ) / 2 =
\Delta_x \, p_x / \hbar$. Here,  $\Delta_x := \sqrt{\hbar / 2 M
\nu_x}$ is the spread in position along the $x$-axis of the
zero-point wavefunction and $p_x$ the corresponding dimensioned
momentum operator. The physics of $H^{p_x}_{\sigma_x} $
cannot be described anymore by Rabi oscillations. In turn,
it yields a conditional displacement in the motion depending on the internal state, producing the so-called
Schr\"odinger cat states~\cite{Solano1,Solano2}. By further
manipulation of laser field directions and phases, we can also
implement $H^{p_y}_{\sigma_y} = 2 \eta_y \Delta_y
\tilde\Omega_y \sigma_y p_y$ and $H^{p_z}_{\sigma_x} = 2
\eta_z \Delta_z \tilde\Omega_z \sigma_x p_z$. This kind of
interactions has already been produced in the lab, under
resonant~\cite{Monroe} and dispersive
conditions~\cite{MonroeNIST}.

We define the wave vector associated with the four ionic internal
levels as
\begin{eqnarray}
| \Psi \rangle := \Psi_a | a \rangle + \Psi_b | b \rangle + \Psi_c
| c \rangle + \Psi_d | d \rangle = \left(
\begin{array}{c}
\Psi_a \\
\Psi_b \\
\Psi_c \\
\Psi_d \\
\end{array}
\right) .  \label{DiracBispinor}
\end{eqnarray}
We may apply simultaneously different laser pulses, with proper
directions and phases, $\eta \equiv \eta_x = \eta_y = \eta_z$,
$\Delta \equiv \Delta_x = \Delta_y = \Delta_z$, $\tilde\Omega
\equiv \tilde\Omega_x = \tilde\Omega_y = \tilde\Omega_z$, $\Omega
\equiv \Omega_x = \Omega_y = \Omega_z$, to compose the following
Hamiltonian acting on $| \Psi \rangle$,
\begin{eqnarray}
H_{\rm D} = \!\!\!\!\!\! && 2 \eta \Delta \tilde\Omega
(\sigma^{ad}_x + \sigma^{bc}_x) p_x + 2 \eta \Delta
\tilde\Omega (\sigma^{ad}_y - \sigma^{bc}_y) p_y \nonumber \\ &&
+ 2 \eta \Delta \tilde\Omega (\sigma^{ac}_x-\sigma^{bd}_x)
p_z + \hbar \Omega (\sigma^{ac}_y+\sigma^{bd}_y) .
\label{DiracIonHamiltonian}
\end{eqnarray}
We rewrite Eq.~(\ref{DiracIonHamiltonian}) in the suitable matrix
form
\begin{eqnarray}
H_{\rm D} \! = \! \left( \begin{array}{cc}  0 &  2
\eta \Delta \tilde\Omega ( \vec{\sigma} \cdot \vec{p} ) \! - \! i \hbar\Omega \\
2 \eta \Delta \tilde\Omega ( \vec{\sigma} \cdot \vec{p} ) \! + \! i \hbar\Omega & 0
\end{array} \right) \! , \label{DiracIonHamiltonianMatrix}
\end{eqnarray}
where each entry represents a $2 \times 2$ matrix. The associated
Schr\"odinger equation, $H_{\rm D} | \Psi \rangle = i \hbar
\partial | \Psi \rangle /
\partial t$, performs the same dynamics as the {\it Dirac
equation} in 3+1 dimensions for a free spin-$1/2$ particle, where
$| \Psi \rangle$ represents the four-component Dirac bispinor.
This is easily seen if we express the Dirac
equation
\begin{equation}
i\hbar\frac{\partial\psi}{\partial t}= {\cal H}_{\rm D} \psi =
(c\vec{\alpha}\cdot\vec{p}+ \beta mc^2)\psi  \label{DiracEquation}
\end{equation}
in its ``supersymmetric'' representation~\cite{Thaller}
\begin{eqnarray}
{\cal H}_{\rm D}  = \left( \begin{array}{cc}  0 & c ( \vec{\sigma} \cdot \vec{p} ) - i m c^2 \\
 c ( \vec{\sigma} \cdot \vec{p} ) + i m c^2 & 0
\end{array} \right) . \label{DiracHamiltonian}
\end{eqnarray}
Here, the $4 \times 4$ matrix $\vec{\alpha}:= ( \alpha_x ,
\alpha_y , \alpha_z ) =
\mbox{off-diag}(\vec{\sigma},\vec{\sigma})$ is the velocity
operator, $\beta:=\mbox{off-diag}(-i\openone_{2},i\openone_{2})$,
and
\begin{eqnarray}
c := 2 \eta \Delta \tilde\Omega \,\,\, , \,\,\,\,\, mc^2 :
= \hbar\Omega , \label{Rosetta}
\end{eqnarray}
are the speed of light and the electron rest energy, respectively.
The notorious analogy between
Eqs.~(\ref{DiracIonHamiltonianMatrix}) and
(\ref{DiracHamiltonian}) shows that the quantum relativistic
evolution of a spin-1/2 particle can be fully reproduced in a
tabletop ion-trap experiment, allowing the study of otherwise
unaccesible physical regimes and effects, as shown below.

In the Dirac formalism, the spin-$1/2$ degree of
freedom is incorporated {\it ab initio}. Moreover, the Dirac bispinor in Eq.~(\ref{DiracBispinor}) is built by components associated with positive and negative energies, $E^{\pm}_{\rm D} = \pm
\sqrt{p^2 c^2 + m^2 c^4}$. This
description is the source of diverse
controversial predictions, as the {\it
Zitterbewegung} and the Klein's paradox.

The {\it Zitterbewegung} is a known quantum-relativistic effect
consisting of a {\it helicoidal motion of a free Dirac particle},
a natural consequence of the non-commutativity of its velocity
operator components, $c \alpha_i$, with $i = x , y , z$. It can be
proved straightforwardly~\cite{Thaller} that the time evolution of the position
operator $\vec{r} = ( x , y , z)$ in the Heisenberg
picture, following $d \vec{r} / dt = \lbrack
\vec{r} , H_{\rm D} \rbrack / i \hbar$, reads
\begin{eqnarray}
\!\!\!\!\! \vec{r}(t) = \!\!\!\!\! && \vec{r} (0) + \frac{4 \eta^2
\Delta^2 \tilde\Omega^2 \vec{p}}{H_{\rm D}} \, t \nonumber \\  &&
+  \left( \vec{\alpha} - \frac{2 \eta \Delta \tilde\Omega
\vec{p}}{H_{\rm D}} \right) \frac{i \hbar \eta \Delta
\tilde\Omega}{H_{\rm D}} \left(e^{2i H_{\rm D} t / \hbar}
- 1\right) . \label{Zitterposition}
\end{eqnarray}
Here, the first two terms on the r.h.s. account for the classical
kinematics of a free particle, while the last oscillating term is
responsible for a transversal ``quivering'' motion. If we consider
a bispinor state with a peaked
momentum around $p_0$, $| \Psi_0 \rangle = | a \rangle
\otimes \exp \lbrack -(p - p_0)^2 / 2 \sigma_p^2 \rbrack$, the
{\it Zitterbewegung}  frequency associated with the measurable
quantity $\langle \vec{r} (t) \rangle$ can be estimated as
\begin{eqnarray}
\omega_{\rm ZB} \approx && \!\!\!\!\! 2 | \bar{E}_{\rm D} | /
\hbar \equiv 2 \sqrt{ 4 \eta^2 \Delta^2  \tilde\Omega^2 p^2_0 / \hbar^2 +
\Omega^2} \nonumber \\ \approx && \!\!\!\!\! 2 \sqrt{ N \eta^2
\tilde\Omega^2 + \Omega^2} ,
\end{eqnarray}
where $\bar{E}_{\rm D} \equiv \langle H_{\rm D} \rangle$ and $N
\equiv \langle a^{\dagger} a \rangle$ are the average energy and
phonon number, respectively. Similarly, we can estimate from
Eq.~(\ref{Zitterposition}) the {\it Zitterbewegung} amplitude
associated with $\langle \vec{r} (t) \rangle$ as
\begin{eqnarray}
R_{\rm ZB}=\frac{\hbar}{2mc}\left(\frac{mc^2}{E}\right)^2=
\frac{\eta \hbar^2 \tilde{\Omega} \Omega \Delta}{4 \eta^2\tilde{\Omega}^2\Delta^2 p_0^2 + \hbar^2 \Omega^2} ,
\end{eqnarray}
and $R_{\rm ZB} \approx \Delta$, if
$\eta\tilde{\Omega} \sim\Omega$.

The standard explanation of
this erratic motion for a free Dirac particle invokes the
interference between the positive- and negative-energy components
of the Dirac bispinor following the dynamics in
Eqs.~(\ref{DiracIonHamiltonianMatrix}) and
(\ref{DiracHamiltonian}). The predicted values for a
real electron, $\omega_{\rm ZB} \sim 10^{21}{\rm Hz}$ and $R_{\rm
ZB} \sim 10^{-3}{\rm \AA}$, are out of experimental reach, the effect has never
been observed, and its existence is even questioned by quantum field theory considerations. To simulate quantum-relativistic effects in other physical systems, like trapped ions or graphene, is not aimed at proving their existence, but at exploiting the differences and analogies in each field. Given the flexibility of trapped-ion systems, we
will have access to a wide range of tunable experimental
parameters~\cite{NIST-Innsbruck}, allowing for realistic and
measurable $\omega_{\rm ZB} \sim 0 - 10^{6}{\rm Hz}$ and $R_{\rm
ZB} \sim 0 - 10^3{\rm \AA}$, depending on the initial vibronic
states.  Due to our piecewise build-up of the 3+1 Dirac
Hamiltonian of Eq.~(\ref{DiracIonHamiltonian}), we can strongly
reduce the experimental demands to study Dirac equations in 1+1
and 2+1 dimensions~\cite{Thaller,ThallerArticle}. In those cases,
the Clifford algebra that characterizes the Dirac matrices is
satisfied by the anticommuting Pauli matrices,
$\{\sigma_i,\sigma_j\}= 2 \delta_{ij}$, where $\{A,B\}:=AB+BA$, given
that we only need 2(3) anticommuting matrices in the 1+1(2+1)
case. Accordingly, the four components of the Dirac bispinor are
conveniently reduced to only two.

We focus now on the 1+1 dimensional case, which could be reached by current experiments, keeping most
striking results available. After a unitary transformation around
the $x$-axis, transforming $\sigma_y$ into $\sigma_z$, the 1+1
Dirac equation stemming from Eq.~(\ref{DiracIonHamiltonian}) can
be cast into $i \hbar \partial | \Psi^{(1)}  \rangle /  {\partial
t} =  H^{(1)}_{\rm D} | \Psi^{(1)} \rangle$ with
\begin{eqnarray}
H^{(1)}_{\rm D} = &&  i \hbar \eta \tilde\Omega \sigma_x
(a_x^{\dagger} - a_x) + \hbar \Omega \sigma_z \nonumber \\ \equiv
&& 2 \eta \Delta \tilde\Omega \sigma_x p_x + \hbar \Omega
\sigma_z , \label{DiracHamiltonian1+1}
\end{eqnarray}
where the Dirac ``spinor'' $| \Psi^{(1)} \rangle = \Psi^{(1)}_a |
a \rangle + \Psi^{(1)}_b | b \rangle$. Note that these two
components are not associated with the spin-$1/2$ degree of
freedom but are linear combinations of positive- and negative-energy
solutions~\cite{ThallerArticle}. Our setup is now reduced to a
simultaneous JC + AJC interaction, $\sigma_xp_x$, and a
Stark-shift term $\sigma_z$ acting on two internal levels. In
Eq.~(\ref{DiracHamiltonian1+1}), the use of $\sigma_z$ is
for the sake of pedagogy, and returning to $\sigma_y$ will not
affect the results.

In the nonrelativistic limit, $m c^2 \gg p_x c$, or equivalently
$\Omega \gg \eta \tilde\Omega$, we may identify the dispersive
limit in Eq.~(\ref{DiracHamiltonian1+1}), and derive the {\it
squeezing} second-order Hamiltonian
\begin{eqnarray}
H^{(1)}_{\rm eff} = \frac{ 2 \eta^2 \Delta^2 \tilde\Omega^2}{\hbar
\Omega} \sigma_z p_x^2 \equiv \sigma_z \frac{p_x^2}{2 m} \,\, ,
\end{eqnarray}
yielding the expected Schr\"odinger Hamiltonian associated with
the classical kinetic energy of a free particle. In the relativistic
limit, $m c^2 \ll p_x c$, which includes $m=0$, the 1+1 case
reduces to $H^{(1)}_{\rm D} = i \hbar \eta \tilde\Omega \sigma_x
(a_x^{\dagger} - a_x)$, which produces Schr\"odinger cats, as
commented above~\cite{Solano1,Solano2}.

For a massless particle, we can show that $d x / dt = \lbrack x ,
H^{(1)}_{\rm D} \rbrack / i \hbar = 2 \eta \Delta \tilde \Omega \sigma_x$
and $d \sigma_x / dt = \lbrack \sigma_x , H^{(1)}_{\rm D} \rbrack
/ i \hbar = 0$. In consequence, $\sigma_x$ is a constant of motion
and the time evolution of the position operator,
\begin{eqnarray}
x(t) =  x(0) + 2 \eta \Delta \tilde\Omega \sigma_x \, t  \,
, \label{ZitterpositionXMass0}
\end{eqnarray}
is classical and does not involve {\it Zitterbewegung}. On the
other hand, for a massive particle, $d \sigma_x / dt = \lbrack
\sigma_x , H_{\rm D} \rbrack / i \hbar = - \Omega \sigma_y$, $d
\sigma_y / dt \neq 0$, and so forth. The produced set of
differential equations yields a {\it Zitterbewegung} oscillatory
solution similar to Eq.~(\ref{Zitterposition}), with the evident
simplification via the replacements of $H_{\rm D}$ by
$H^{(1)}_{\rm D}$, $\vec{p}$ by $p_x$, and $\vec{\alpha}$ by
$\sigma_x$. It is noteworthy to mention that
the phenomenon of mass acquisition, which could be done here in a continuous manner by raising the coupling strength $\Omega$, is related to the spontaneous
symmetry breaking mechanism of the Higgs field.

At this stage, it is clear that the measurement of the expectation
value of the position operator, $\langle x(t) \rangle$, as a
function of the interaction time $t$ is of importance. A recent
proposal for realizing fast measurements of motional
quadratures~\cite{Solano3} relies on the possibility of measuring
the population of an ionic internal level, $P_a$, at short
probe-motion interaction times $\tau$ with high precision~\cite{Meekhof}. Given
an initial vibronic state $\rho (0) = | +_{\phi} \rangle \langle
+_{\phi} | \otimes \rho_{\rm m}$, where $| \pm_{\phi} \rangle = (
| a \rangle \pm e^{i \phi} | b \rangle ) / \sqrt{2}$ and
$\rho_{\rm m}$ describes an unknown motional density operator,  we
can make use of
\begin{eqnarray}
\langle Y_{\phi} \rangle =\frac{d}{d \tau} P_{\rm e}^{+_{\phi}}
(\tau) \bigg\vert_{\tau = 0} , \label{MeasurePosition}
\end{eqnarray}
where the generalized quadrature $Y_{\phi} = ( a e^{-i \phi} -
a^{\dagger} e^{i \phi} ) / 2i$. Then, position and momentum
operators are measured when choosing $\phi = - \pi / 2$ and $\phi
= 0$, respectively. To apply this technique, it is required a
particular initial state of the internal states, so the
measurement of $\langle x(t) \rangle$ would have to be done in two
parts, one associated with each of the suitably projected states,
$| +_{\phi} \rangle$ and $| -_{\phi} \rangle$.

We turn now to the possible simulation of Klein's paradox. In
1929, Klein noticed \cite{Klein1} the anomalous behavior of Dirac
particles in regions where a high potential energy $V$ exists:
$H^{(1)}_V = H^{(1)}_{\rm D} + V \openone_2$. When $V > 2 m c^2$,
negative-energy electrons (components) may swallow $V$, acquiring
positive energy and behaving
 as ordinary electrons, while leaving a hole in the
Dirac sea. This stems from the fact that the relativistic energy
related to $H^{(1)}_V$ may be recast into $(pc)^2 = (E^{(1)}_{\rm
V} - V +  m c^2)(E^{(1)}_{\rm V} - V - m c^2)$, which is positive
when either both factors are positive or negative.  In the second
case, the total energy $E^{(1)}_{\rm V} < - m c^2 + V$ can
be larger than $m c^2$, as noticed by Klein. In this case an
$e^--e^+$ (electron-positron) pair could be created from $V$. The
sudden raise of the constant potential $V (|a\rangle \langle a | +
|b\rangle \langle b | )$, at a certain time $t = t_0$ after an
evolution associated with $H^{(1)}_{\rm D}$, could simulate this
phenomenon. We suggest to produce potential $V$ with the required
characteristics through a fast and homogeneous Stark shift in both
internal levels. The natural way to detect Klein's paradox,
assuming an initial positive-energy internal state $| a \rangle$ ($p_0 = 0$),
is via measurement of nonzero population in the
negative-energy component $| b \rangle$.

Finally, in as much as massless chiral fermions in condensed matter, we could also produce a 1+1 axial
anomaly~\cite{Nielsen} by changing Eq.~(\ref{DiracHamiltonian1+1}) into
$H_D^{(1)}=c\sigma_xp_x+qEx$, where $x  \propto (a + a^{\dagger})$ is a motional displacement.

The basic ingredients to implement the 1D Dirac dynamics of
Eq.~(\ref{DiracHamiltonian1+1}) with a single trapped ion are two
independent electronic (internal) states coupled to a
one-dimensional motional degree of freedom. The required states
could be composed by two ground-state hyperfine levels of an earth
alkaline atomic ion, e.g. of $^{25}$Mg$^+$ by $| F =3;m_f = -3\rangle$ and
$|F =2;m_f = -2\rangle$, $|a\rangle$ and $|b\rangle$, respectively,
as depicted in Fig.~1. A constant external magnetic field will
define the quantization axis and lift the degeneracy of  levels
being potentially useful to provide additional states (like
$|c\rangle$ and $|d\rangle$ in Fig.~1) necessary for higher
dimensional simulations, see Eq.~(\ref{DiracIonHamiltonian}).

\begin{figure}
\vspace*{-.5cm}\hspace*{0.5cm} \epsfig{file=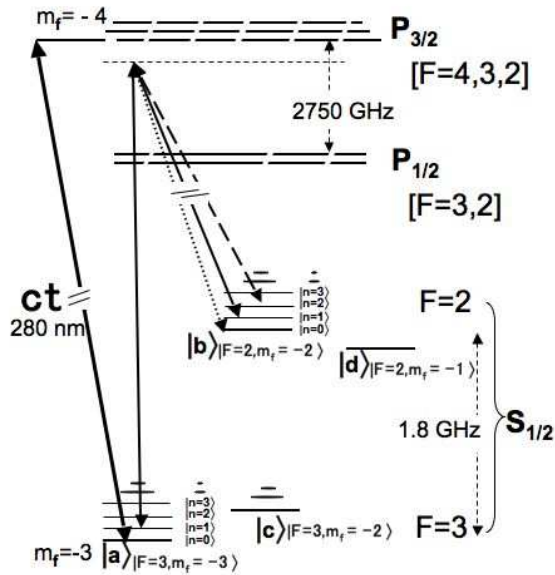,width=12cm} \caption{Relevant
energy levels (not to scale) of a $^{25}$Mg$^+$ ion. Shown are the
ground-state hyperfine levels supplying the two internal states, $|a\rangle$ and $|b\rangle$, and the equidistant harmonic
oscillator levels related to the harmonic axial confinement in a
trap similar to that described in~\cite{rowe02,blatt03}. We
subsumed excited levels of the $P_{1/2}$ and $P_{3/2}$ states. We depict
the resonant transition state sensitive detection and the relevant
types of two photon stimulated  Raman transitions between states
$|a,n=1\rangle$ and $|b,n=0,1,2\rangle$: red sideband (dotted line), carrier (solid line), and blue sideband (dashed line).}
\end{figure}

At the start, the ions will be laser cooled
close to the motional ground state and optically pumped into state
$|a\rangle$~\cite{king}. One red/blue sideband and one carrier
transition will be driven simultaneously~\cite{winelandroyal,Monroe}
to implement the desired dynamics of Eq.~(\ref{DiracHamiltonian})
via two-photon stimulated Raman transitions. To measure the ion position we rely on the mapping of motional information on the internal degrees of freedom~\cite{Solano3,Meekhof}
and take advantage of the high fidelity of state sensitive
detection realized by an additional laser beam, tuned to a cycling
transition~\cite{monroegate}, coupling state $|a\rangle$
resonantly to the $P_{3/2}$ level. Considering the available laser intensities~\cite{axel06},
all necessary Raman beams could be derived from a single laser
source. We split the original beam and provide the necessary
frequency offsets, phase control and switching via multi-passing
through Acusto-Optical-Modulators~\cite{winelandroyal}.The number of laser beams could be further reduced by
electro optical modulators to provide red and blue sidebands
simultaneously~\cite{MonroeNIST,Monroe}. To implement an overall shift in
the potential (Stark shift of level $|a\rangle$ and $|b\rangle$ in Klein's paradox)
without changing their mutual energy difference we chose the
directions and polarizations of the Raman beams appropriately.

In conclusion, we have shown how to simulate the Dirac equation in
3+1 dimensions for a free spin-$1/2$ particle, and
quantum-relativistic effects like the {\it Zitterbewegung} and
Klein's paradox, in a single trapped ion. We have studied the 1+1
case, where experimental needs are minimal while keeping most
striking predictions. We believe that these simulations open
attractive avenues and a fruitful dialog between different
scientific communities.

The authors thank D. Leibfried and J. I. Latorre for valuable comments. L.L. acknowledges support from MEC FPU grant No. AP2003-0014. L.L. and J.L. were partially supported by the Spanish MEC FIS2005-05304 and CSIC 2004 5 0E 271 projects. T.S. acknowledges support by DFG, MPQ, and MPG. E.S. is grateful for the hospitality at CSIC (Madrid) and acknowledges support from EU EuroSQIP and DFG SFB 631 projects.

\end{document}